\documentclass[letterpaper]{article}
\usepackage{aaai23} 
\usepackage{times} 
\usepackage{helvet} 
\usepackage{courier} 
\usepackage[hyphens]{url} 
\usepackage{graphicx} 
\usepackage{graphicx} 
\usepackage{natbib} 
\usepackage{caption} 
\usepackage[export]{adjustbox}
\usepackage{booktabs}
\usepackage{xcolor}
\usepackage{soul}
\usepackage{amsmath}

\usepackage{comment}
\excludecomment{mycomment}

\title{Collective Intelligence in Human-AI Teams: A Bayesian Theory of Mind Approach}

\author {
    Samuel Westby\textsuperscript{\rm 1},
    Christoph Riedl\textsuperscript{\rm 2},
}
\affiliations {
    \textsuperscript{\rm 1} Network Science Institute, Northeastern University, Boston, MA\\
    \textsuperscript{\rm 2} Khoury College of Computer Sciences, Northeastern University, Boston, MA\\ 
    westby.s@northeastern.edu, c.riedl@northeastern.edu
}

\begin{document}
\maketitle
\begin{abstract}
We develop a network of Bayesian agents that collectively model the mental states of teammates from the observed communication. Using a generative computational approach to cognition, we make two contributions. First, we show that our agent could generate interventions that improve the collective intelligence of a human-AI team beyond what humans alone would achieve. Second, we develop a real-time measure of human's theory of mind ability and test theories about human cognition. We use data collected from an online experiment in which 145 individuals in 29 human-only teams of five communicate through a chat-based system to solve a cognitive task. We find that humans (a) struggle to fully integrate information from teammates into their decisions, especially when communication load is high, and (b) have cognitive biases which lead them to underweight certain useful, but ambiguous, information. Our theory of mind ability measure predicts both individual- and team-level performance. Observing teams' first 25\% of messages explains about 8\% of the variation in final team performance, a 170\% improvement compared to the current state of the art.
\end{abstract}

\section{Introduction}

The reliance on teamwork in organizations \cite{wuchty2007increasing}, coupled with remarkable recent progress in artificial intelligence, have supercharged the vision to develop collaborative Human-AI teams \cite{malone2015handbook,oneill2020human}. 
Human-AI teams promise to overcome human biases and information processing limitations, reaching performance higher than human-only teams could \cite{brynjolfsson2018artificial}.
Despite some recent advances \cite[e.g.,][]{bansal2019updates, pynadath2022disaster, seraj2022learning} there remain significant difficulties in developing agents that interact with multiple, heterogeneous humans working on cognitive tasks engaged in cooperative communication in an ad-hoc team. Here, we draw on research of cognitive processes to develop Human-AI teams and explain collaborative decision making. 

To communicate efficiently, humans infer the beliefs, opinions, knowledge, and related states of mind of other people \cite{nickerson1999hidden,call2011does}. This is referred to as social perceptiveness or theory of mind \cite[ToM;][]{premack1978does}. Recent research on collective intelligence, has provided a wide range of empirical evidence suggesting that ToM (and related processes governing collective memory, attention, and reasoning) is a significant predictor of human collective intelligence \cite{woolley2010evidence,riedl2021quantifying,woolley2022collective,engel2014reading}. 
Indeed, ToM is especially beneficial for interdependent cognitive tasks that benefit when teams leverage their members’ expertise \cite{lewis2003measuring}. 
Some work suggests that ToM is the mechanism that allows collectives to use more performance-relevant information from their environment than a single individual without such connections could, for example, by facilitating a balance between diversity and cognitive efficiency \cite{riedl2017bursty,hong2004groups}. 
As human civilization shifts further toward knowledge work \cite{autor2014skills} where the most value is realized if members fully use and integrate their unique expertise, this ability is increasingly important.


\begin{figure*}[t!]
    \centering
    \includegraphics[width=1\linewidth]{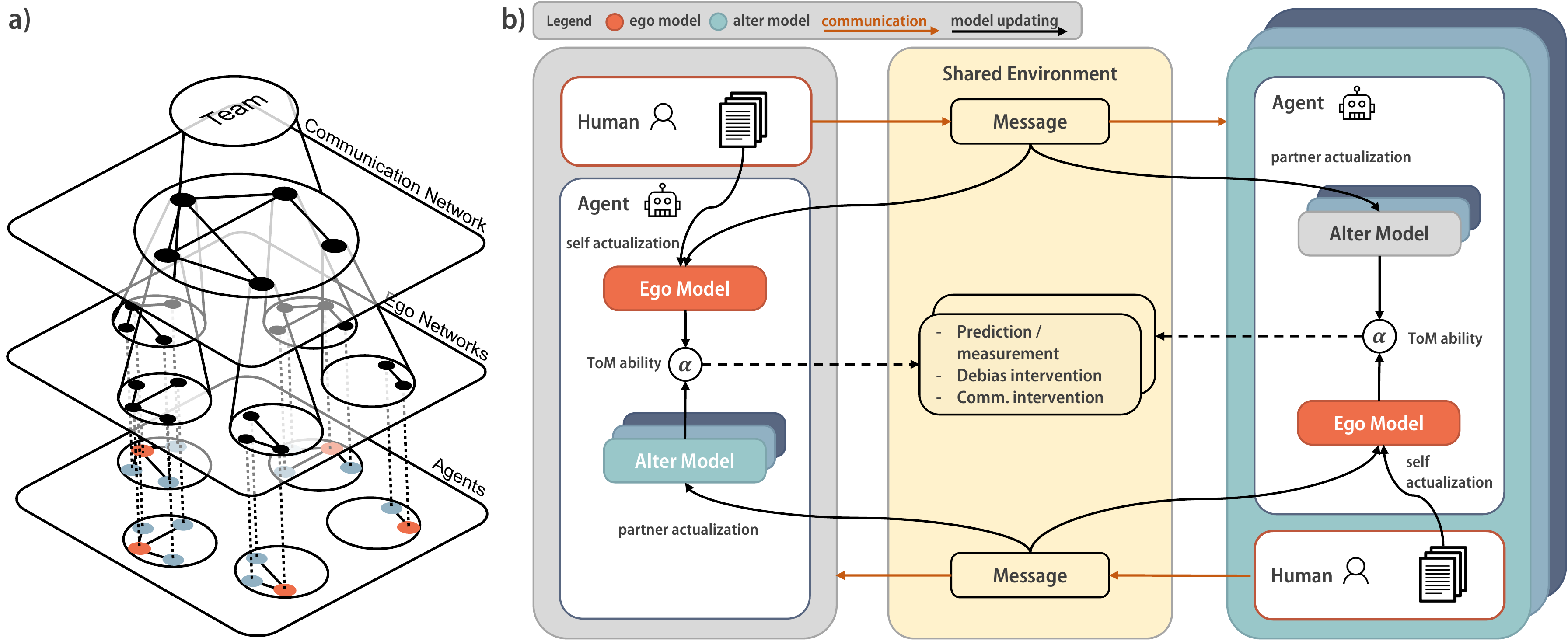}
    \caption{\textit{Framework of human-AI teaming with Theory of Mind (ToM).} 
        a) Nested layers of ToM agents. Agents model ego networks of the individual they shadow. 
        b) Every human team member is paired with an AI agent. Humans send messages to others through a shared environment. The ToM agent infers beliefs for both own ideas (\textit{Ego Model}), and ideas of others (one \textit{Alter Model} per network neighbor). \textit{Ego Model} is updated with initial information and new knowledge generated by the human through self actualization. \textit{Alter Models} are updated based on incoming messages from teammates through partner actualization. Agents combine information from the ego and alter models with weighting determined $\alpha$ denoting ToM ability.
    }
    \label{fig:conceptual}
\end{figure*}

Recent work has started to develop a formal account of collective intelligence to explain the relationship between individual interaction and collective performance using (approximate or variational) Bayesian inference \cite[or free energy;][]{friston2010free,friston2013life,heins2022spin}. The free energy principle is a mathematical framework for multiscale behavioral processes that suggests a system of self-similar agents self-organizes by minimizing variational free energy in its exchanges with the environment \cite[Fig.~\ref{fig:conceptual}a;][]{friston2006free}. Recent extensions have applied the framework to explain human communication \cite{vasil2020world}. A key advantage of this approach is that free energy minimization can be translated into a generative, agent-based process theory \cite{friston2017active,kaufmann2021active}. 
This generative theory provides a computational approach to cognition \cite{tenenbaum2011grow,griffiths2015manifesto} that allows us to simultaneously (a) build agents for Human-AI teams that are highly explainable but also (b) test theories about human cognitive processes and measure human theory of mind \textit{ability} in real time. This promises to advance our understanding of a key process of human collective intelligence. The current state of the art to measure theory of mind---the Reading the Mind in the Eyes test \cite{baron2001reading}---is a static, indirect, survey-based instrument, which typically explains about 3\% of the variation \cite{riedl2021quantifying}. 

In this paper, we develop a Bayesian theory of mind agent that can form ad-hoc mental models about its teammates, based exclusively on observations drawn from human communication (Fig.~\ref{fig:conceptual}b). We use data collected from a large, IRB approved human-subject experiment in which 145 individuals in 29 teams of five, randomly assigned to different social networks controlling the team's communication topology, communicate through a chat-based system to solve a Hidden Profile task \cite{stasser1985pooling}. 
We then simulate artificial teams in which each human is shadowed by an agent. The agent observes the same incoming and outgoing messages as the human did in the experiment. Modeling human behavior with our generative AI model allows us to test whether people do indeed form mental models of what their teammates know, how effectively they do so, and whether this ability is predictive of team performance. 
In a last step, we perform a counterfactual simulation to demonstrate how our Bayesian agent could trigger effective interventions that would increase Human-AI team performance over the observed human-only teams. 

Our work provides a framework that expands theory of mind (and collective intelligence more broadly) from a static construct to a dynamical one that may vary according to situational factors, for example, due to changes in arousal, anxiety, and motivation with dynamically changing task requirements, time pressure, and recognition \cite{qin2022predictive,balietti2021incentives}. We contribute to a body of research that has so far mostly used toy models---often using only a single agent---with an application to real data from multi-human communication \cite{vasil2020world,kaufmann2021active,albarracin2022epistemic,heins2022spin}. Our work generates important cognitive insights into how humans communicate and reason to uncover hidden profiles. In summary, we make four main contributions.


\begin{enumerate}
    \item We develop a networked Bayesian agent that models beliefs using human-human communication. We apply the agent to data from a human team experiment and demonstrate how the agent can monitor theory of mind in real time, predict both correct and human answers, and intervene to raise human performance. 
    \item We find the model accurately captures the decisions made by humans, varying in predictable ways with experimental stimuli like network position and task difficulty. Comparing model fits with simpler ``lesioned'' ToM models shows the value contributed by each component. 
    \item We develop two real-time measures for human theory of mind \textit{ability}. The first, based on observed human communication and decisions, explains 51\% variation in final team performance. The second, based on communication alone, explains 8\% variation in \textit{final} team performance after observing just the first quarter of communication, a 170\% improvement compared to the current state of the art, the Reading the Mind in the Eyes test. 
    Simulations of artificial human-AI teams suggest a significant 4\% performance increase from AI triggered interventions.
    \item We contribute to cognitive theory by presenting empirical evidence that cognitive biases explain the shortfall of human performance, such as a tendency to under-weight ambiguous information and failure to fully integrate information provided by others. We explain temporal patterns showing that high functioning teams send the most useful information early before converging on common beliefs.
\end{enumerate}


\section{Related Work}
\paragraph{Human-Agent teaming.}

A long history of Human-Agent teaming has evolved alongside technological developments. Early examples such as Vannevar Bush's Memex \cite{bush1945we} demonstrate a longstanding fascination with augmenting human performance. Recently, work has specialized in many sub-fields including understanding mental models and related constructs like team situational awareness \cite{chen2014human,converse1993shared, glikson2020human}. For example, effective Human-AI interaction has been shown to rely critically on the ability to form mental models about what AI teammates are doing \cite{bansal2019updates,paleja2021utility,bansal2019beyond,gero2020mental,alipour2021improving}.

Significantly less work has focused on how AI can form ``mental models'' of humans. \citet{fugener2022cognitive} highlight this disparity by identifying situations where humans having mental models of the AI are not helpful, while AI having ``mental models'' of humans are. 
Given the challenges of designing multi-agent systems, human-AI teaming work has often focused on studying pairs of one agent and one human  \cite[e.g.,][]{bansal2019beyond,bansal2019updates,baker2011bayesian,fugener2022cognitive,alipour2021improving}. 
Furthermore, past work has often side-stepped challenges posed by language-based communication 
by constraining the scope to spatial or highly stylized tasks \cite{kaufmann2021active,baker2017rational,khalvati2019modeling}. Others use Wizard of Oz techniques \cite{schelble2022let,hohenstein2022vero} to facilitate communication-based human-AI teaming interaction. To build an autonomous AI teammate that improves human-only team performance, one must build agents that overcome these obstacles. This becomes more challenging in a team of unfamiliar teammates, without a priori knowledge, while learning dynamically from language-based communication \cite{stone2010adhoc}.


\paragraph{Multi-agent Bayesian models}

Multi-agent Bayesian models have been used to study coordination \cite{khalvati2019modeling, wu2021too}, opinion dynamics \cite{albarracin2022epistemic}, efficient information fusion \cite{pavlin2010multi}, and theory of mind \cite{baker2017rational}. This can be modeled as a partially observable Markov decision process (POMDP) for each agent where the states are the set of other agents' beliefs and observations are dependent on other agents' actions \cite{smith2022step}.

\section{Hidden Profile \& Human Subject Data}
A primary advantage of teams over lone individuals when solving complex problems is their ability to expand the pool of available information, thereby enabling teams to reach higher quality solutions  \cite{mesmer2009information}. The Hidden Profile task \cite{stasser1985pooling} is a research task designed to mimic this decision-making scenario in which individuals hold private knowledge \cite{stone2010adhoc}. In the task, some information is commonly held among all team members while each individual is also endowed with unique private information. Subjects do not know what information is shared or private. Information sharing is the central process through which teammates collectively solve the task \cite{mesmer2009information}; conversely, failing to share all available information causes them to come to incorrect conclusions. 
Despite the importance of information sharing for team performance, past research has shown teams often deviate from the optimal use of information \cite{stasser1985pooling}. Discussions tend to reinforce information held in common, rather than share information held uniquely by one team member \cite{nickerson1999hidden}. One reason for this is that individuals impute their own knowledge on others and hence assume that private information is already shared \cite{nickerson1999hidden}.

This gives rise to the ``hidden profile'' and directly points to avenues in which AI may improve team performance: identifying which information is uniquely held by each teammate and encouraging them to share it. Specifically, an agent may detect if their own mental model diverges from the inferred mental model of another (``I know something that I believe you don't know'') indicating a window of opportunity for an effective intervention. 
It also provides the basis for our measure of theory of mind \textit{ability}. Individuals who form more precise mental models of their teammates (and who impute less of their own knowledge on others) will be more efficient communicators who share more useful information in a more targeted manner.

We use data from an IRB approved online experiment conducted on the Volunteer Science platform \cite{radford2016volunteer} in which 145 individuals in 29 human-only teams of five solved a Hidden Profile task (data and code available at \url{https://github.com/riedlc/HumanAITeamsAndCI}). The task is framed as a crime investigation: the team needs to pool clues to answer questions about the target, culprit, and time of an art heist. There are six clues for each question. When combined, the correct answer out of five possible options is obvious. One randomly selected clue is given to every teammate (the public clue) and each individual receives one of the remaining five clues (the private clue, also randomly selected). Teams were randomly assigned to a communication topology using all 21 possible connected five node graphs (e.g., star, ring, chain; see Fig.~\ref{fig:conceptual}a for one example). Teams then communicate via text-based chat where each message is sent to all neighboring teammates. After a five minute discussion phase, each individual submits answers for the culprit's identity, the target, and the day of the art heist. Subjects were recruited from Amazon Mechanical Turk \cite{paolacci2010running} and paid a \$0.75 flat fee for participation as well as a \$0.25 performance bonus for each correct answer. The entire task took about seven minutes to complete. Subjects are blind to the network topology they have been assigned to as well as the total number of individuals on their team. For simplicity of exposition and analysis, we rely only on the \textit{culprit} dimension of the task. We compute individual performance as $\{0, 1\}$ depending on whether the culprit guess is correct and team performance as the average individual performance (majority voting does not substantively change results). On average, individuals received $3.1$ (SD $1.9$) chat messages from each partner. 

To make initial clues and communication machine interpretable, we manually code the content as \textit{strong no (SN), maybe no (MN), maybe yes (MY)}, and \textit{strong yes (SY)} for each of the five answer options (the inferred states). This creates a set of 20 possible observations. We translate messages into likelihoods for the inferred states using fixed values, either estimated from the data using maximum likelihood estimation (MLE) from a grid search 
or using untrained intuitive values. For example, the message,``it might be \#4'', would be coded as \textit{maybe yes} with a likelihood of $1.4$ for $\#3$, leaving the likelihoods for the states not mentioned in the message unaffected. Ambiguous statements and messages related to team coordination were coded as ``neutral'' and dropped from the analysis. Notice that agents form beliefs solely based on the observed human communication, even if humans make certain statements about wrong facts (e.g., ``strong yes \#4'' when the correct answer is \#3), or ambiguous statements about correct facts. Agents can thus form wrong beliefs \cite{albarracin2022epistemic}.


        

\section{Bayesian Multi-Agent Model}


We create a networked Bayesian agent for each individual. Each agent ``shadows'' one human, observing the same messages  (both are inside the same Markov blanket; Fig.~\ref{fig:conceptual}a) and infers human beliefs and derive the answer to the Hidden Profile task. 
That is, our Bayesian system is a model of beliefs about hidden states of their environment. Ideally, the state inferred after the communication phase is identical to the correct answer to the Hidden Profile task. The resulting model has five parameters: four information weights \textit{SN, MN, MY, SY} determining the likelihood distribution of observations under inferred beliefs, and the theory of mind ability $\alpha_D$ which modulates the relative weighting of the self vs.~partner beliefs which we describe in more detail below.

\begin{figure}[t!]
    \centering
    \includegraphics[width=1\linewidth]{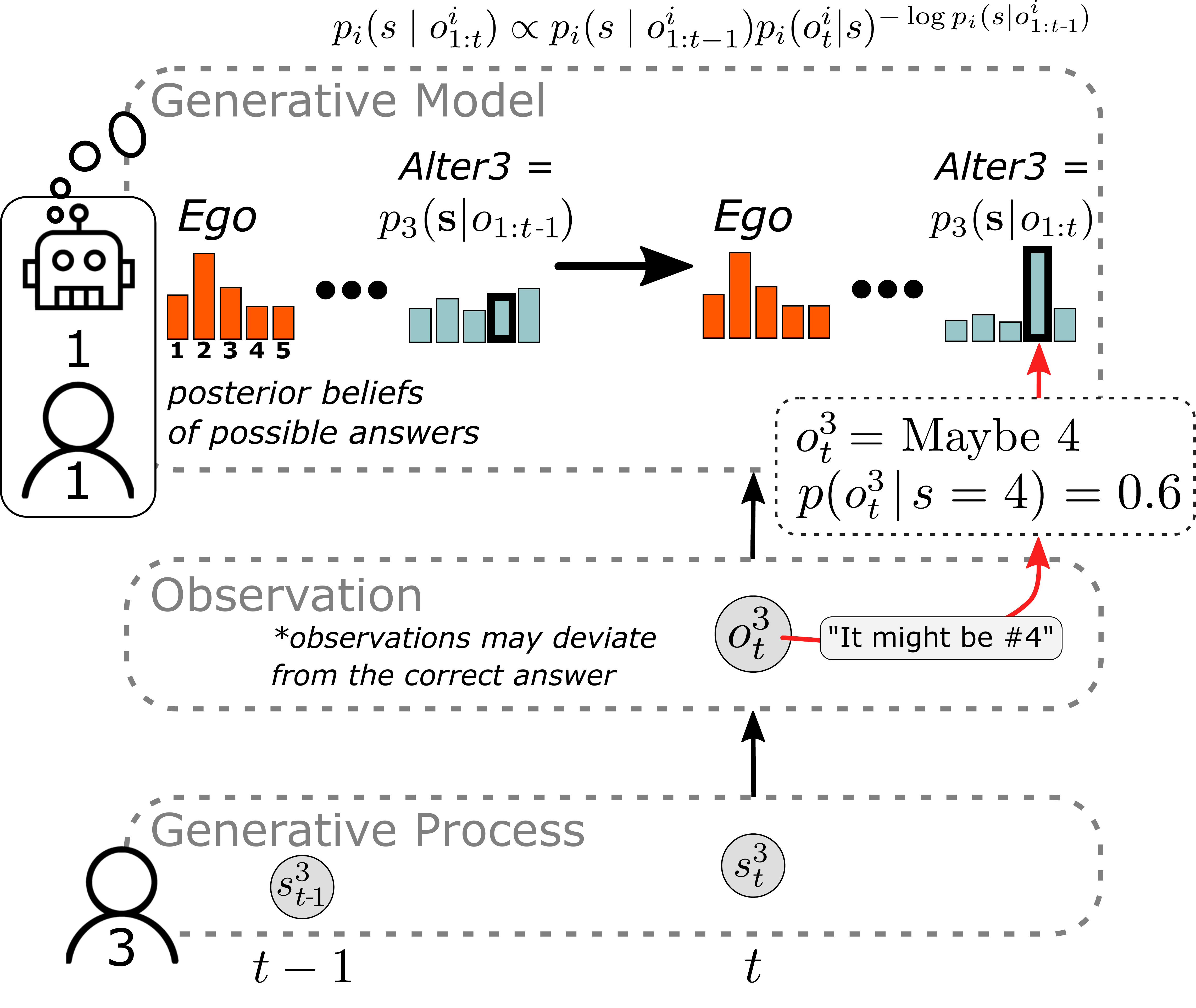}
    
    \caption{\textit{Bayesian model of Theory of Mind.} 
        Agent 1, shadowing Player 1, models all teammates in Player 1's ego network (Markov blanket). In Agent 1's generative model, \textit{Alter3} corresponds to Agent 1's beliefs of Player 3's beliefs. At time $t$ Player 3 says, ``It might be \#4'', which is coded as a \textit{MY} for answer 4. Given this new observation, Agent 1 uses Equation ~\ref{eq:update} to update its beliefs about Player 3's beliefs.
    }
    \label{fig:model}
\end{figure}

\paragraph{Mental Models.}
We use Bayesian inference to infer a posterior distribution $p(\textbf{s} \mid \textbf{o})$ over states $\textbf{s}$ (five answer options), from a set of observations $\textbf{o}$ (messages sent between players in the recorded chat). Since there are five discrete states, we can compute posteriors directly without the need to approximate them. More complicated environments may require the use of approximate inference methods like 
free energy/active inference \cite{friston2006free}. 

Agents are comprised of one \textit{Ego Model} and one \textit{Alter Model} for each neighbor (Fig.~\ref{fig:conceptual}b). That is, the model follows a multi-agent paradigm 
with independent mental models nested within an agent. All models hold a posterior distribution of inferred states, but differ in how they are initialized and updated. \textit{Ego Models} are initialized with priors derived from the public and private clues assigned to the player (paper stack icons in Fig.~\ref{fig:conceptual}b) and updated with \textit{outgoing} messages from a player (\textit{self actualization}). \textit{Alter Models} are initialized with uniform priors and updated with \textit{incoming} messages of the corresponding (\textit{partner actualization}). 



Mental models are updated by accounting for the surprise of the observation. Surprise-weighting thus encodes a preference for posteriors with low surprise. That is, the effect of new observations is diminished relative to the \textit{Ego} or \textit{Alter} model's existing posterior using 
\vspace{-.5cm}
\begin{equation}
\label{eq:update}
p_i(s \mid o^i_{1:t}) \propto p_i(s \mid o^i_{1:t-1}) p_i(o^i_t | s) ^{\overbrace{\scriptstyle-\log{p_i(s \mid o^i_{1:t-1})}}^\text{surprise}}
\end{equation}
where $s$ is a state and $o^i_t$ is an observation (message) sent by player $i$ at time $t$. The likelihood is raised to the negative log of the previous time step's posterior. 

\begin{table*}[h!]
    \centering
    \begin{tabular}{lccclc}
        \toprule
             & Performance & Human-Agent &  &  \multicolumn{1}{c}{Model Comparison} \\
             & (\% Correct) & Agreement & LogLik &  \multicolumn{1}{c}{(Likelihood Ratio test)}  &$\alpha_D$\\
        \midrule
        \textbf{Human}           & \textbf{66.2\%}    \\
        Random          & 19.6 $\pm$ 3.1\%    & 20.0 $\pm$ 3.0\% & -231.759 \\
        Prior only     & 48.8 $\pm$ 3.1\%    & 46.6 $\pm$ 2.9\% & -215.906 & vs.~Random $p < 0.0001$ \\
        ToM (self-actualization only, MLE)      & 63.6 $\pm$ 1.8\% & 73.0 $\pm$ 1.9\% & -146.372            & vs.~Prior-only $p < 0.0001$ & 0\\
        ToM (partner-actualization only, MLE) & 76.7 $\pm$ 1.0\% & 66.3 $\pm$ 1.0\% & -133.653 & vs.~Self-only $p < 0.0001$ & 1\\
        ToM (MLE)                               & 72.8 $\pm$ 0.8\% & 75.1 $\pm$ 1.0\% & \textbf{-106.640}   & vs.~Partner-only $p < 0.0001$ & 0.95\\
        ToM (max performance)                   & \textbf{77.2 $\pm$ 0.8\%} & 71.3 $\pm$ 0.9\% & -109.826   & vs.~MLE $p = 0.012$ & 0.95\\
        ToM (max agreement)                     & 71.8 $\pm$ 0.7\% & \textbf{79.7 $\pm$ 0.8\%} & -118.464    & vs.~MLE  $p < 0.0001$ & 0.45\\
        \midrule
        With random intervention                & 79.0 $\pm$ 1.8\% & 70.8 $\pm$ 1.6\% &&& 0.95 \\
        With intervention                       & \textbf{82.1} $\pm$ 0.7\% & 70.0 $\pm$ 0.6\% && vs.~{Rand. int}. $p < 0.0001$ & 0.95\\
        \bottomrule
    \end{tabular}
    \caption{
        \textit{Model evaluation results and comparison with human behavior.} 
        P-values based on likelihood ratio test. We calculate standard deviations over 100 trials. Information weights (\textit{SN, MN, SY, MY}) are learned from the data through a grid search: Prior (0.05, 0.05, 2, 2), self-act.~(0.05, 0.05, 1.5, 2), partner-act.~(0.15, 1, 1.55, 2), MLE~(0.1, 1, 1.45, 2), max perf.~(0.35, 0.85, 1.95, 2), max agg.~(0.05, 0.75, 1.25, 1.95). Interventions use the same parameters as max performance and $t$-test for comparison.
    }
    \label{tab:results_table}
\end{table*}



\paragraph{Agent.}
The agent is the hierarchical coordinator and aggregator of its mental models. The ToM ability parameter $\alpha_D$ modulates the relative weight with which the agent combines its \textit{Ego} and \textit{Alter} models.
We conceptualize $\alpha_D$ as the ability to accurately infer beliefs of other agents and pay attention to them \cite{apperly2009humans}. It represents the relative weighting between the agent's own \textit{Ego Model} and its \textit{Alter Models}. When $\alpha_D=0$, the \textit{Alter} posterior is uniform and has no effect and the final prediction is based only on the \textit{Ego Model}. When $\alpha_D=1$, the \textit{Alter Models} are weighted equally to the \textit{Ego Model}. 
Agent $i$ aggregates its mental models into a final posterior distribution using
\begin{equation}
\label{eq:aggregate}
 p_i(s \mid \textbf{M}_i) \propto p_i\left(s \mid \textbf{o}^i\right) \prod_{\substack{m \in \textbf{M}_i\\m \neq i}} p_m\left(s \mid \textbf{o}^m\right)^{\alpha_D}
\end{equation}
where $\textbf{M}_i$ is Agent $i$'s set of mental models and $p_m\left(s \mid \textbf{o}^m\right)$ is posterior of state $s$ for the mental model of player $m$ over $\textbf{o}^m$, the set of $i$'s observations of $m$.

\paragraph{Alternative Models.}
To test whether the full representational capacity of theory of mind with both \textit{self-actualization} and \textit{partner-actualization} loops are necessary to understand human mental states, we formulate two alternative models that ``lesion'' one or both of the updating loops. This allows us to test whether it is possible to explain human inferences about their teammates without appealing to a fully developed theory of mind. We compute $p$-values from likelihood ratio tests comparing the models.

\section{Results}

\begin{figure*}[t!]
    \centering
    \includegraphics[width=.329\linewidth]{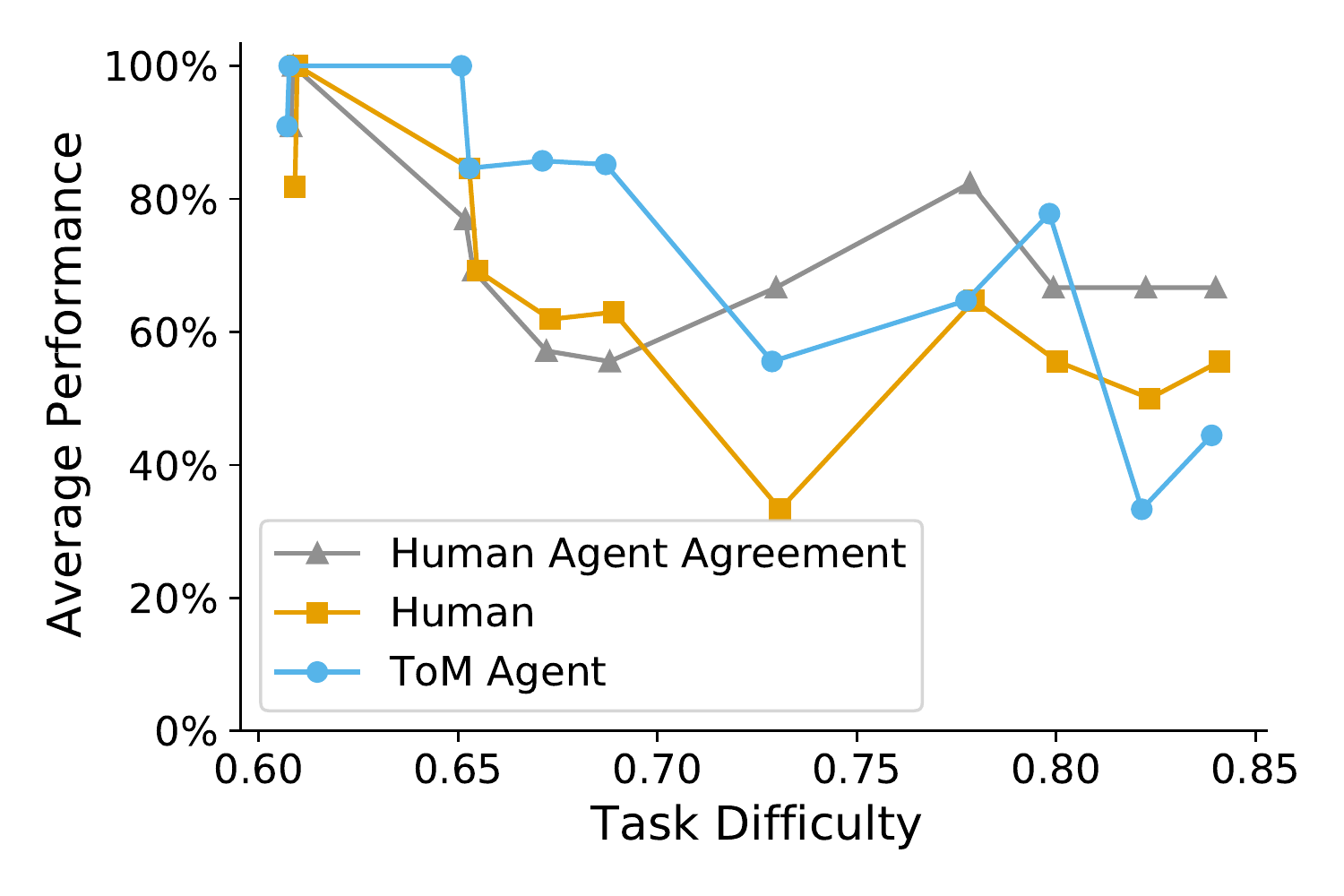}
    \includegraphics[width=.329\linewidth]{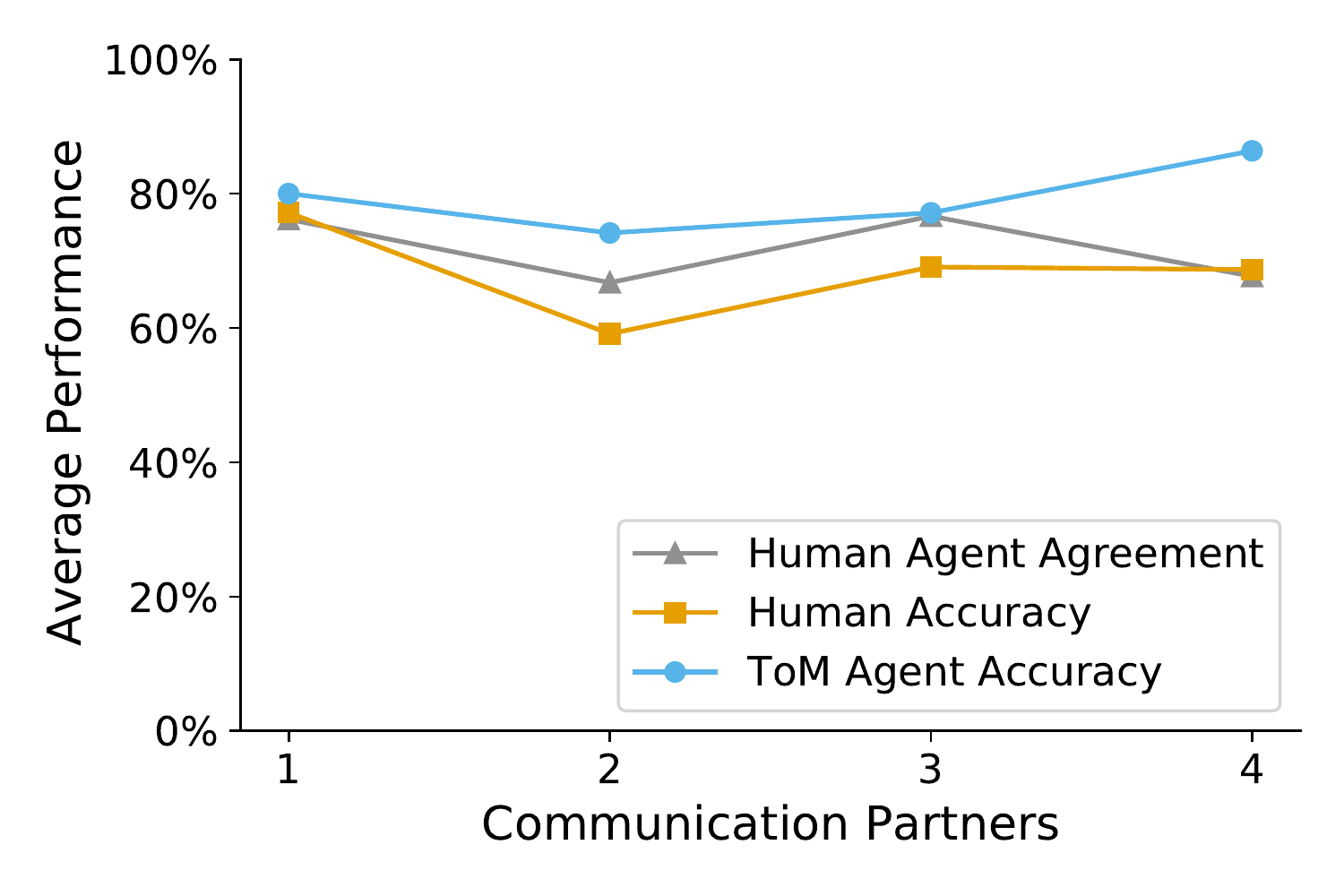}
    \includegraphics[width=.329\linewidth]{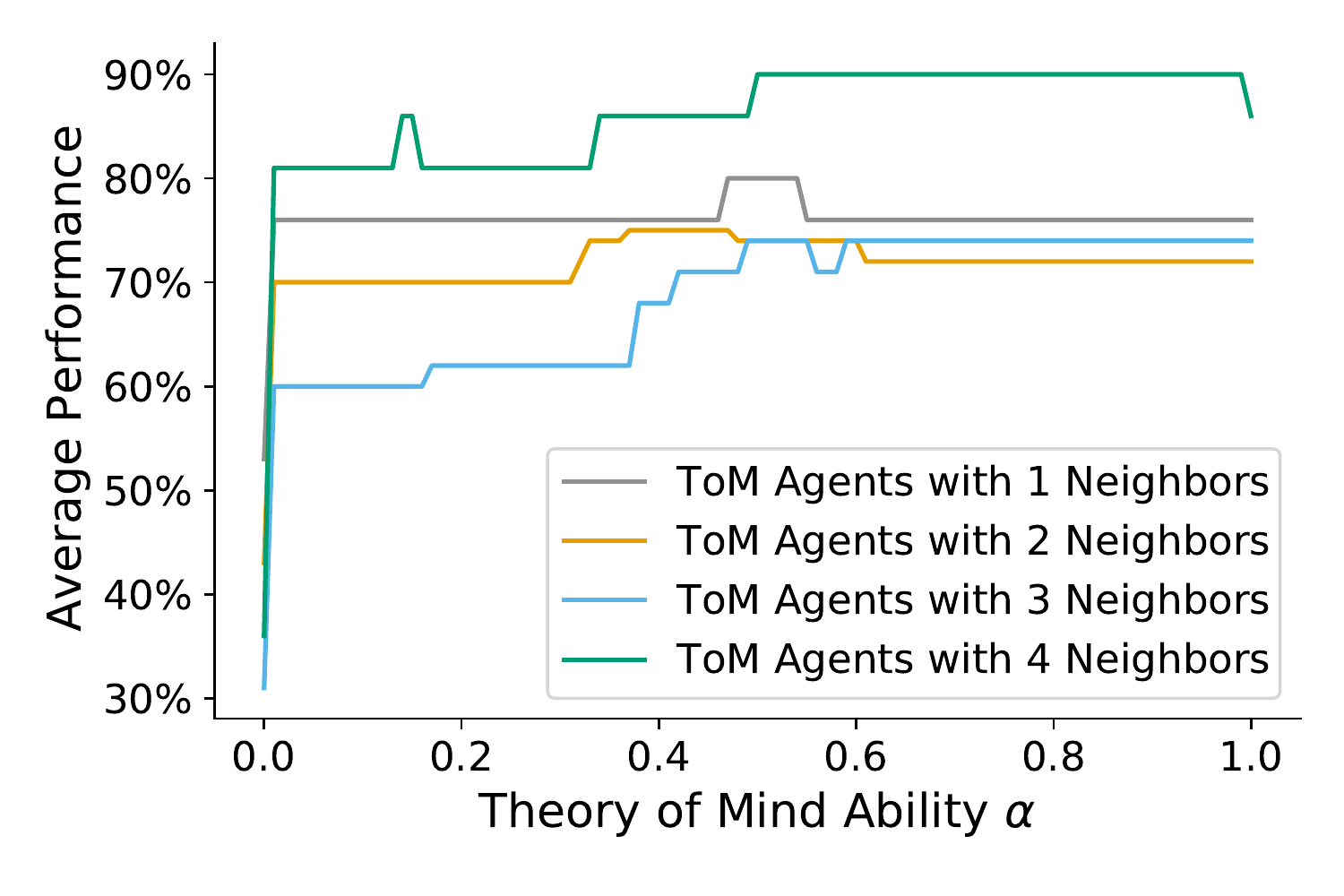}
    
    \caption{\textit{Human performance varies with task difficulty and number of communication partners.} From left to right:
        a) Human performance decreases with task difficulty and is outperformed by AI agent in most cases.
        b) Agent improves over human performance especially when communicating with many teammates.
        c) Agents with many communication partners benefit most from high ToM ability $\alpha_D$.
        (parameters: $\alpha_D=0.95$, $SN=0.35; MN=0.85; MY=1.95; SY=2$).
    }
    \label{fig:results_panel}
\end{figure*}

\paragraph{Model Evaluation.}
We find strong support for the hypothesis that humans use Bayesian inference to model the minds of their teammates and communicate and make decisions according to those models (Table ~\ref{tab:results_table}). Compared to a model using only prior information (the clues distributed in the experiment), a model capturing humans' ability to update their own beliefs (self-actualization only) fits the data significantly better. A model allowing humans to update beliefs about their teammates (partner-actualization only) fits significantly better still. Finally, a model including the capability to update both own and partner beliefs has the highest fit. 
Higher values for $\alpha_D$ generally lead to more peaked posterior distributions. 
This explains why the parameter values that produce the highest likelihood differ slightly from those of the highest accuracy ($\alpha_{D}^{\mathit{MLE}}=0.95$ vs.~$\alpha_{D}^{\mathit{max acc}}=0.45$). In summary, the comparative fit analysis provides reliable evidence for the mutually inferred alignment of attention (cf., mental states) among teammates.

Our model accurately captures the judgments of human participants, varying in predictable ways with random experimental manipulation of task difficulty and the number of communication partners. 
We measure the task difficulty faced by each individual based on how much information the individual can draw about the correct answer from the two clues they initially received. This captures how difficult it is for an individual to guess the correct answer before communicating with other players (this is a somewhat noisy measure as it ignores team-level effects of the clue distributions). 
Not surprisingly, human performance decreases with task difficulty, suggesting that humans suffer from cognitive overload (Fig.~\ref{fig:results_panel}a). Our agent achieves high accuracy predicting human's incorrect answers under high task difficulty (high true-negative rate). 

Human performance varies with the number of communication partners (Fig.~\ref{fig:results_panel}b). Given the nature of the task, access to more communication partners should be beneficial as this guarantees access to more information. Humans, however, perform worse with more communication partners while our ToM agent achieves its highest performance when placed in the most central network position (agent is 20\% better than human with four partners). This suggests that humans struggle to integrate information when communicating with many teammates. 
This picture becomes even clearer when contrasting this with ToM ability $\alpha_D$ (Fig.~\ref{fig:results_panel}c). Higher levels of ToM ability $\alpha_D$ have the highest benefit on performance in central network positions, yet $\alpha_D$ hardly matters when connected to just a single teammate.

\paragraph{Analysis of Human Decision Biases.}
The ToM model predicts with high accuracy instances in which humans provide the correct answer as well as those in which they provide the wrong answer (48\% true-negative accuracy). Comparing the information weighting parameters for optimal performance with those for the highest model fit with data from the human subject experiment from the MLE estimates, we can directly see why human performance falls short. Humans pay not enough attention to information ruling out alternatives (optimal information weighting for \textit{strong no} $0.25$ vs.~MLE fit $0.05$). The difference is even more pronounced for ambiguous information (optimal information weighting for \textit{maybe no} $0.9$ vs.~MLE fit $0.05$): humans undervalue information that is ambiguous, yet crucial in arriving at the correct answer. Because this information is ambiguous, humans may attempt to make sense of it by imputing their own understanding (i.e., resorting to their own prior) instead of updating their beliefs in the direction of the ambiguous message.
A similar weighting difference for \textit{maybe yes} statements suggests that humans communicate \textit{strong yes} information in vague ways (maybe-ing their statements) and could significantly improve their performance by placing higher weight on such statements (or communicating them more forcefully). 


\paragraph{Measuring Theory of Mind.}
We propose two measures of human theory of mind ability: $\alpha_D$ and $\alpha_C$. The first, $\alpha_D$, is based on an individual's ability to form and integrate accurate mental models of others when making \textit{decisions} and corresponds directly to our model parameter that governs the relative weighting of the \textit{Ego} vs.~\textit{Alter Models}. The second, $\alpha_C$, captures an individual's ability to \textit{communicate} the most useful information. We perform maximum likelihood estimate using a grid search over the relevant parameter space \cite{balietti2021optimal}. Then, we fix the maximum likelihood estimate of the nuisance parameters for information weighting (\textit{SN, MN, MY, SY}) but consider the marginal of all values of $\alpha_D$. Instead of then picking the global best fitting value for the entire data set, we pick the maximum likelihood estimate of $\alpha_D$ separately for each individual. That is, we use the model's inner ToM working to estimate which value of individual $i's$ $\alpha_D$ produces the highest likelihood of the observed decision. For the second measure $\alpha_C$, we consider outgoing messages sent by each individual and compute the expected surprise that this message should produce for the recipient, relative to ego's \textit{Alter Model} of the recipient. Notice that we compute this internally within the Markov blanket of an agent. We do not use information about how surprising the message is for the recipient but rather how useful the sender thinks it should be relative to what they think the recipient knows. Intuitively, individuals who possess a high theory of mind ability, will be better at sending the right message to the right person compared to those with lower ToM ability. Both measures capture social perceptiveness: how much attention an individual pays to what others in the team know. 

\begin{figure*}[t!]
    \includegraphics[width=.329\linewidth,valign=t]{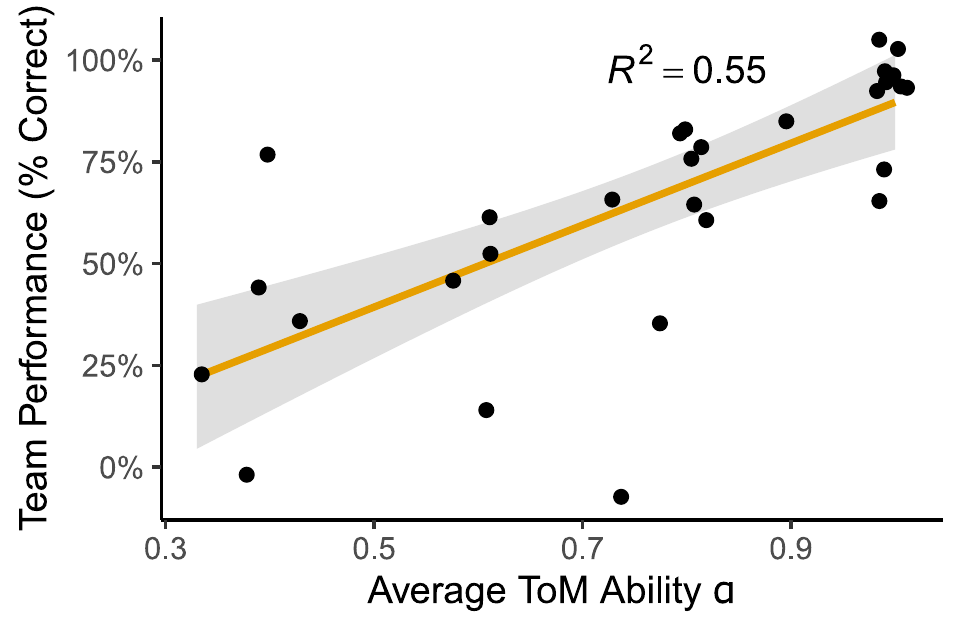}
    \includegraphics[width=.329\linewidth,valign=t]{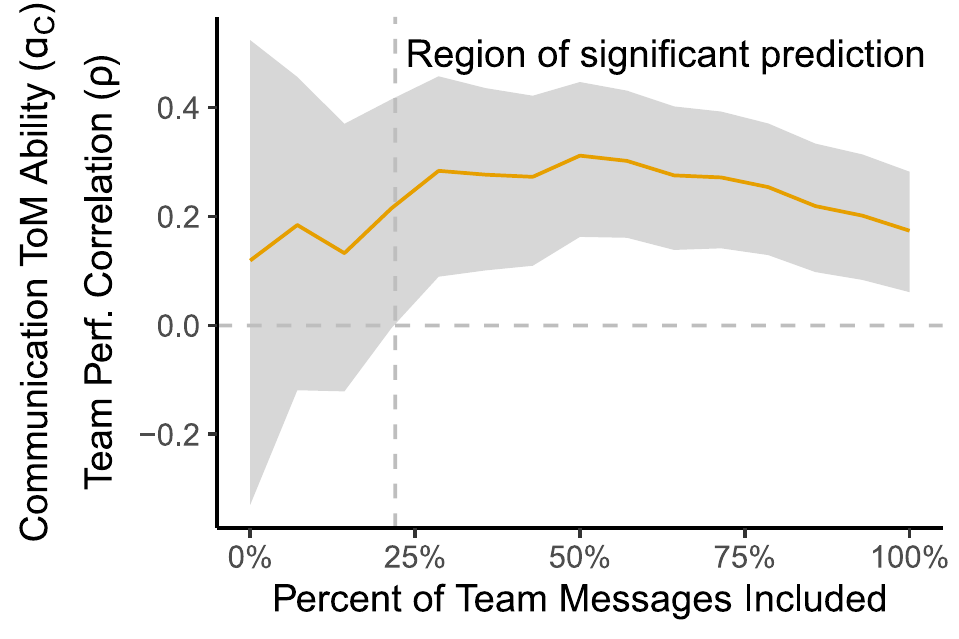}
    \includegraphics[width=.329\linewidth,valign=t]{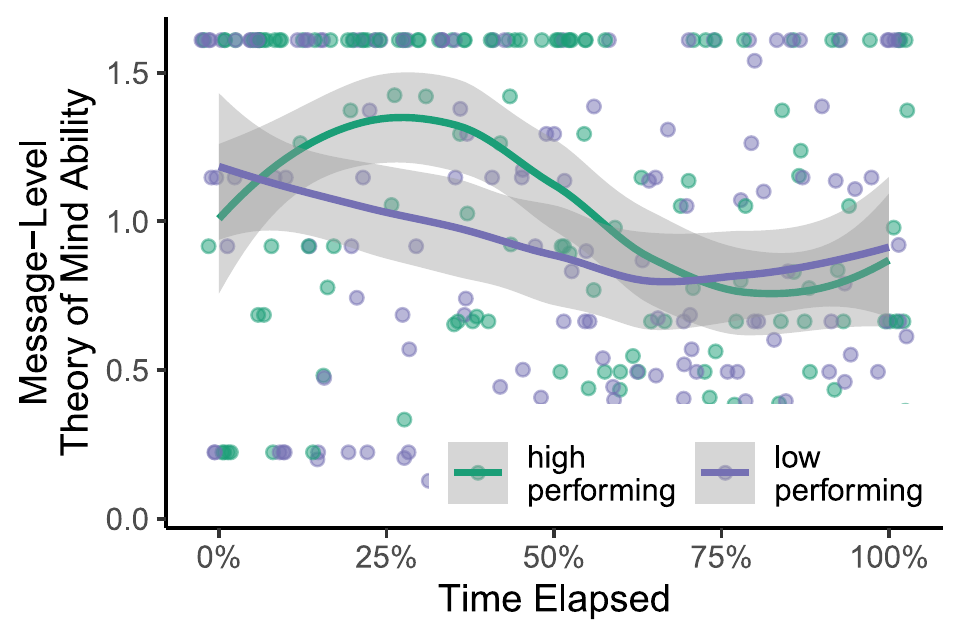} \\
    \caption{\textit{Theory of Mind ability predicts team performance.}
        a) Team average ToM $\alpha_D^\mathit{team}$ is a strong predictor of the final team performance.
        b) Communication ToM $\alpha_D^\mathit{team}$ serves as a real-time measure of collective intelligence. Only about the first 25\% of team messages are necessary to make significant predictions of \textit{final} team performance.
        c) High- and low-performing teams have markedly different temporal patterns of ToM $\alpha_C$. 
    }
    \label{fig:humanPerformance}
\end{figure*}

We find that individual-level ToM ability $\alpha_D$ is a strong predictor of individual-level performance ($\beta = 0.59; p < 0.001; R^2 = 0.26$). 
Aggregating to the team level, we find that average ToM ability $\alpha_D^\mathit{team}$ is a strong predictor of final team performance (Fig.~\ref{fig:humanPerformance}a). We find that the effect of ToM ability is moderated by average betweenness centrality suggesting team performance increases most when high-ToM ability $\alpha_D$ individuals occupy high betweenness network positions ($\beta = 0.39; p = 0.04$). The amount of communication sent within a team, notably, is \textit{not} a significant predictor of team performance ($\beta = -0.00; p = 0.265$).

Turning to our analysis of theory of mind communication ability $\alpha_C^\mathit{team}$, we find that it is  a strong predictor of team-level performance ($\beta = 0.47; p = 0.019$). Given that we can measure $\alpha_C$ on the message level, it can serve as a real-time measure of theory of mind. We find that after observing only the \textit{first} 25\% of a team's messages, $\alpha_C^\mathit{team}$ is a significant predictor of \textit{final} team performance (Fig.~\ref{fig:humanPerformance}b). We analyze the temporal pattern in which high- vs.~low-performing teams communicate (Fig.~\ref{fig:humanPerformance}c). High-performing teams send messages with high information content (high surprise) early during the team task but then send consolidating, low information content messages at the end to facilitate convergence (low surprise). We illustrate this in the example below ($^*$'s indicate high surprise messages with novel content).


\begin{quote}
Human 1: \textit{It will not happen on Tuesday}$^*$\\
Human 3: \textit{No Wednesday or Friday}$^*$\\
Human 2: \textit{Monday or Thursday}$^*$\\
Human 2: \textit{Did you get a no Thursday info?}\\
Human 5: \textit{Yeah I got no Thursday}$^*$\\
Human 3: \textit{I got no Thursday}\\
Human 5: \textit{So it must be Monday}
\end{quote}

This suggests that team cognition is not static but instead emerges dynamically and that high-performing teams have the collective capacity to modulate shared cognition dynamically to achieve both efficient information transfer and convergence during different periods of the task. 
 Low-performing teams on the other hand fail to send high information content messages and also fail to achieve convergence sending more surprising messages late during the task. This pattern illustrates that high information content alone is not desirable. Instead, convergence and joint attention to team consensus are crucial \cite{woolley2022collective}.

The current standard to predict social perceptiveness is the Reading the Mind in the Eyes (RME) test \cite{baron2001reading,almaatouq2021task}. Using data from a large meta analysis \cite{riedl2021quantifying} of 5,279 individuals in 1,356 groups, we find RME explains between 0\% and 3\% of the variation in team performance (depending on the task). Our $\alpha_C^\mathit{team}$ measure explains 8\% after observing only 25\% of team communication, an improvement of about 170\%. Our proposed measure captures social perceptiveness passively and in real time which can be used to interpret a team’s current status and determine opportunities for interventions. Furthermore, RME captures social perceptiveness of a single individual, while our measure is group based. Our work also extends previous measures of ``information diversity'' \cite{riedl2017bursty}. It thus captures aspects of collective attention and memory \cite{gupta2020emergence}.

\paragraph{Human-Agent Team Performance.}
So far, our AI agent was only passively shadowing its assigned human, reasoning about the mental states of that human and its connected teammates. In this section, we extend this reasoning to allow the agent to trigger interventions that could be deployed in human-AI teams and quantify what performance improvement this might yield. 
We perform a counterfactual simulation in which we allow each AI agent to identify and send one message to each network neighbor. Each agent compares its \textit{Ego Model} against its \textit{Alter Models} to identify divergence in inferred beliefs. Each agent then draws from the set of messages it received during the team discussion and chooses a message to send to each neighbor.
To do this, the agent calculates the effect of sharing one of its available messages on the \textit{Alter Models} and shares the message that results in the lowest KL divergence, defined as
$D_\text{KL}(Q\, ||\, P) = \sum_i Q(i) \ln \frac{Q(i)}{P(i)}$,
between the \textit{Ego} and \textit{Alter}$_i$ posteriors over all five possible answer options. If no message lowers the KL divergence, the agent shares no message. This is summarized as taking the action $a_{ij}$ for each agent $i$ and neighbor $j$ where 
\begin{equation}
\label{eq:action}
a_{ij} = \underset{m \in \textbf{o}}{\textit{arg\,min}} \, D_\text{KL}\left(p_{\text{ego}_i}(\textbf{s } | \textbf{ o})\, || \, p_{\text{alter}_j}(\textbf{s } | \textbf{ o'}, m)\right)
\end{equation}
$m$ is selected from the set of messages \textbf{o} that agent $i$ sent or received, \textbf{s} is a vector of the five possible answers, and \textbf{o'} is the set of messages agent $i$ received from agent $j$. To establish a baseline intervention, we let $a_{ij}$ be a random message in $\textbf{o} \cup \{\text{no message\}}$. Here, performance improves to 79.0 $\pm$ 1.8\% averaged over 100 trials. For the targeted intervention, performance improves  4.9\% to 82.1 $\pm$ 0.7\% which is significantly higher ($t$-test $p < 0.0001$) than the random intervention.
Notice that this intervention would not be possible without our ToM-based multi-agent model. Without it, we could not determine which message to send to which alter.

\section{Discussion}
We develop a framework that combines theory of mind, Bayesian inference, and collective intelligence into a generative computational model. Our model accurately captures the decisions made by human participants in a cognitive decision-making experiment, varying in predictable ways with experimental manipulation of task difficulty and network position. 
Our results suggest that humans use Bayesian inference and Theory of Mind to model their own beliefs and those of their teammates and communicate and make decisions according to those models. We provide empirical evidence that humans do not do this perfectly but suffer from cognitive biases. Nonetheless, our Bayesian agent is robust and achieves high performance even when fed biased and incorrect information, providing a pathway to implement high-performing human-AI teams. Notably, our agent works in ad hoc teams with heterogeneous partners without any pretraining. In such human-AI teams, our AI could augment humans' limited cognitive memory, attention, and reasoning abilities to increase collective intelligence.

We show empirical evidence that the collective dynamics of Bayesian agents updating probabilities of hypotheses using observations, collectively predict the performance at the team level. This provides the basis for a real-time measure of theory of mind ability, and maybe even collective intelligence more broadly \cite{heins2022spin}. The better the mental models of the team members align---the less surprising observations drawn from communication become---the higher the team’s collective intelligence. Our implementation of direct surprise weighting could be extended with a fuller implementation of the free energy principle that would allow agents to learn asymmetric beliefs about the reliability of their partners' signals.
Taken together, this is a framework to capture the emergence of collective memory, attention, and reasoning in real time \cite{luria1973working,gupta2020emergence}.


\begin{mycomment}


Individually-held cognitive understanding of team members’ skills, knowledge and beliefs are a form of emergent social cognition - a shared system of encoding, storing and retrieving information across distributed across agents that is updated based on communication via message passing \cite{lewis2003measuring}. 





The Bayesian theory of mind approach to human-agent teams outperforms humans in the Hidden Profile task and leads to an estimation of theory of mind ability. It lays a foundation for future work where modeled mental states inform human-agent team interventions and advance theory of collective intelligence.

Our work goes beyond prior research that merely sensing the content of one member’s work and proactively alerting them to related information (e.g., Gimpel et al., 2020).

our system automatically builds in calibration: it won’t generate tons of messages for everyone but it’s targeted: send the required information to the exact right person.

Our ToM communication ability measure using surprise could be improved by using a fuller free energy model to better capture the positive benefits of convergence (sending low surprise messages at the end). Specifically, we theorize that team level free energy functions as a real-time predictor of collective intelligence. 

ToM explains how / why info propagates through the network. If someone invents new info it needs to be shared because other person doesn’t know yet

This work helps answer important questions about the emergence of collective intelligence 
Research challenge to find good process measures \cite{riedl2021quantifying}.  collective attention and memory \cite{gupta2020emergence}.

\end{mycomment}

\cleardoublepage
\section*{Acknowledgements} 
This work was supported by the Army Research Laboratory [Grant W911NF-19-2-0135].

\bibliography{manuscript}

\end{document}